\documentclass[sigconf, manuscript]{acmart}

\AtBeginDocument{%
  \providecommand\BibTeX{{%
    \normalfont B\kern-0.5em{\scshape i\kern-0.25em b}\kern-0.8em\TeX}}}

\copyrightyear{2021}
\acmYear{2021}
\setcopyright{acmlicensed}\acmConference[IMX '21]{ACM International Conference on Interactive Media Experiences}{June 21--23, 2021}{Virtual Event, NY, USA}
\acmBooktitle{ACM International Conference on Interactive Media Experiences (IMX '21), June 21--23, 2021, Virtual Event, NY, USA}
\acmPrice{15.00}
\acmDOI{10.1145/3452918.3458806}
\acmISBN{978-1-4503-8389-9/21/06}

\graphicspath{{figures/}{pictures/}{images/}{./}}
\newcommand{\etal}{\textit{et al}.}
\definecolor{greensz}{rgb}{0.0, 0.5, 0.0}

\usepackage{url}
\usepackage{outlines}
\usepackage{caption}
\usepackage{subcaption}
 \usepackage{csquotes}
\colorlet{red}{black}

\begin{document}

\title[3D Sketching and Speech for Interactive Model Retrieval in VR]
{Mixing Modalities of 3D Sketching and Speech for Interactive Model Retrieval in Virtual Reality}

\author{Daniele Giunchi}
\affiliation{\institution{University College London} \country{United Kingdom}}

\author{Alejandro Sztrajman}
\affiliation{\institution{University College London} \country{United Kingdom}}

\author{Stuart James}
\affiliation{\institution{Visual Geometry and Modelling (VGM) Lab, Istituto Italiano di Tecnologia}\country{Italy}}

\author{Anthony Steed}
\affiliation{\institution{University College London} \country{United Kingdom}}

\renewcommand{\shortauthors}{Trovato and Tobin, et al.}

\begin{abstract}
Sketch and speech are intuitive interaction methods that convey complementary information and have been independently used for 3D model retrieval in virtual environments. While sketch has been shown to be an effective retrieval method, not all collections are easily navigable using this modality alone. We design a new challenging database for sketch comprised of 3D chairs where each of the components (arms, legs, seat, back) are independently colored. To overcome this, we implement a multimodal interface for querying 3D model databases within a virtual environment. We base the sketch on the state-of-the-art for 3D Sketch Retrieval, and use a Wizard-of-Oz style experiment to process the voice input. In this way, we avoid the complexities of natural language processing which frequently requires fine-tuning to be robust. We conduct two user studies and show that hybrid search strategies emerge from the combination of interactions, fostering the advantages provided by both modalities.

\end{abstract}




\maketitle

\section{Introduction}
In recent years we have witnessed a rise of consumer interest for Virtual Reality (VR), largely motivated by improved performances and lowering commercial prices of devices~\cite{berg2016,oculus,vive}. This has led to new opportunities for content development for artistic or design purposes~\cite{bekele2018,tiltbrush,virtualist,BIMx,oneirosvr}. Such tasks often involve finding a 2D/3D asset in a collection, which has been achieved in the past using input modalities such as speech, gestures and sketching.
Within VR, these modalities have been shown to be more intuitive than traditional desktop interfaces relying on mouse and keyboard inputs~\cite{boves2002must,wahlster2006smartkom, van2009follow}.
They help convey complementary information to identify a target in a collection, although their combination in multimodal inputs usually requires a delicate weighting to be effective.
The advent of deep learning algorithms has brought new methods that can learn to combine multiple modalities such as speech, gestures, eye-gazing, and sketches~\cite{10.1145/3107990.3107996}. However, such additional inputs demand the development of more complex algorithms and analysis techniques to be interpreted effectively.
Sketches are iconicized representations of objects, and many variables affect its interpretability, including the user's drawing skills, educational level, mood and utilized tools~\cite{eitz2012hdhso}. Sketch-based Image and Video Retrieval has received much attention since sketching is intuitive~\cite{bui2018sketching}, depicts visual information very quickly and can be easily carried out in 2D using a mouse or touch screen (optionally with a pen), additionally supporting semantic information provided through UI tools (drop-downs or search)~\cite{james2014interactive} for video.
In contrast, 3D sketch as query-by-example is an under-explored research area in the context of virtual environments. Recent works have shown that 3D sketch alone does not lead to high accuracy results~\cite{10.1145/3359997.3365751}. Giunchi~\etal~\cite{giunchi20183d} address this issue by presenting the user with a navigable set of query results instead of a single one. Furthermore, they implement a generalization of the retrieval task, achieving similar results with different object categories. 
Nevertheless, we will show that the sketch-based search is insufficient to enable the effective navigation of databases \textcolor{red}{with high variation of colors among objects parts. Sketches carry only visual information, not a semantic one, such as the meaning of a stroke representing a part of a chair.} 
This imposes important limitations in a search query, forcing the user to focus exclusively on similarities related to shape.
Hence, we design a new dataset of chairs built from ShapeNet~\cite{shapeNet}, by segmenting each chair into four parts and coloring them from a selection of six colours. We show that this simple addition already provides a granularity level to the database, which makes sketch-based search unfit for efficient retrieval.
We choose chairs for our dataset due to their relative simplicity. As drawing a chair relies largely on primitive concepts (lines), this allows us to disentangle user drawing-skills during our studies.
In order to overcome the limitations of sketch-based retrieval, we implement a multimodal system that extends previous works on VR sketching applications.
We leverage the recent advances of immersive 3D sketch model retrieval based on deep learning models~\cite{giunchi20183d} and create a generic pipeline where 3D sketch and speech interactions are integrated and their feature descriptors can be interchanged seamlessly.
In our design, the user is immersed in a virtual environment and interacts via both sketch and speech to search for a target object from our database. 
Thus, the expressive power of the queries is extended beyond the visual domain by introducing semantic descriptions for shapes and colors.
%
To evaluate our system, we design a user study to compare the three different interaction modalities: only sketch, only speech, and sketch plus speech combined. We analyze the participants' interaction via individual modes, and we study the emergence of new strategies when they are allowed to combine them.
All three experimental configurations allow incremental search, with the possibility of step-by-step refining. Before starting our study, we ran a preliminary experiment to determine the optimal number of words-per-queries for the speech interaction for our user test.
The contributions of this paper are as follows:
\begin{itemize}
    \item We create a large variational database of segmented chairs, and we explore how 3D sketch-based retrieval alone performs in such collection. We discover that it is unfit for an efficient navigation.
    \item We design a multimodal interface for 3D model retrieval in VR with both sketch and voice inputs. As part of our design, we implement a consistent translation method between both types of queries, allowing their integration during a single search session.
    \item We perform a user study to evaluate our multimodal interface's search performance and the corresponding single modes of interaction. We focus our analysis on the emergence of new search strategies, enabled by the integration of voice and sketch interactions during the study.
\end{itemize}
In Section 2, we give an overview of related work on virtual environments and multimodal interaction. Section 3 explains the details of our virtual environment model retrieval system and the novel use of interactive machine-learning-based searches that enable an iterative sketch and query refinement process. Section 4 presents a preliminary user study to determine the optimal number of words-per-query for our speech interaction, and perform general fine-adjustments to our experimental design. In Section 5, we study the emergence of new strategies for search refinement during the user studies and analyze the user performance in terms of accuracy, speed and user experience rating. Section 6 presents the limitations of our design and ideas for future work. In Section 7, we present our conclusions.


\section{Related Works}
\label{sec:related}


We first review independent modalities of sketch (sec.~\ref{sec:SketchInVE}) and voice (sec.~\ref{sec:speechinvirtualenvironment}) interactions within a virtual environment. Finally, we describe how previous studies of the combination of modalities, especially sketch and voice, influenced our work (sec.~\ref{sec:multimodalInteraction}).

\subsection{Sketch in Mixed Environment}  \label{sec:SketchInVE}
In the last few years, sketch interaction has inspired many studies, especially in combination with 3D environments. In VRSketchIn~\cite{Drey2020}, Drey~\etal~use a 6DoF-tracked pen and tablet to explore 3D mid-air sketches in combination with a 2D constrained sketch on a surface. They explore 2D and 3D sketch in design space, analyzing these two different metaphors and related emerging patterns.
Elsayed~\etal~\cite{Elsayed2020} showcase VRSketchPen that is a pen that provides unconstrained 3D sketching by using two haptics modalities. The first one is a pneumatic force able to replicate a real surface constraint, and the second one is vibro-tactile feedback for emulating textures. 
A different approach is described by Gasques~\etal~\cite{Gasques2019} with PintAR. The user is provided with a pen, a tablet and an AR device, and the system is able to port the sketch taken on the tablet in the augmented environment. Additional functionalities are added such as rotation, translation, a sketch library and customizable views. While this study exploits the 2D sketch interaction moving the content in an AR space, Leiva~\etal~\cite{Leiva2020} designed Pronto, a video prototyping system that uses a tablet and merges 3D manipulation with 2D video. In particular, Pronto supports four modes:
3D spatial data capture, 2D sketch depicting and positioning in 3D environment and animation generation. With these simple operations, Pronto produces a video that shows the prototype of an augmented reality experience.
Object design benefited a lot from augmented reality applications, such as in Reipschlager~\etal's DesignAR~\cite{Reipschlager2019}. This system includes an AR workstation with the possibility of navigating through a database of AR objects, sketching with a pen in a 2D surface, extruding contours. The integration of pen and multi-touch improve the accuracy that usually represents an issue for mid-air sketching systems.
Suzuki~\etal~\cite{Suzuki2020} propose RealitySketch, an alternative approach of drawing sketches that are moved into the augmented world. The user sketches and connects the drawing to a real object exploiting real dynamics and physical phenomena in their prototype. For example, the user can sketch a line that connects a ball and a pivot point and the arc that determines the angle of the pendulum, and the sketched item will follow the expected pendulum oscillation.
Other works try to use 3D sketch as primary interaction to retrieve models comparing traits and colors to shapes and textures. Li~\etal's~\cite{Li:2016:SSR:3056462.3056474} work compares multiple 3D sketch-based algorithms for 3D model retrieval. They gather hand-drawn 3D sketches and generate a dataset paired with a 3D model collection. Giunchi~\etal~\cite{Giunchi:2018:SIM, 10.1145/3359997.3365751, 8446609} implemented a VR application that uses sketch as the main mechanism to search a target chair among an extensive collection. This software provides the user with an iterative method of depicting a chair, querying a neural network with a 3D sketch at each step, or by fusing the sketch and a chair model. 
We follow the same implementation of this sketching system and insert the additional speech interaction component in this work.

\subsection{Speech in Virtual Environments}
\label{sec:speechinvirtualenvironment}
Natural language interfaces have been widely used in many different areas such as databases~\cite{10.1007/978-3-642-82815-76, Li:2014:NIN:2588555.2594519}, 
mobile systems~\cite{gruenstein-etal-2008-multimodal, 10.1007/0-387-23152-837}, home media systems~\cite{10.1007/978-3-319-60366-723, 10.1007/978-981-13-5758-917} 
and vehicle interfaces~\cite{Bernsen01exploringnatural, Kikuchi1992CARFOLLOWINGMB}.
In the case of VR, the addition of voice input can increase the sense of embodiment and the interface's intuitiveness. The design of a voice interface requires tackling a variety of tasks~\cite{mcglashan1995speech}: speech recognition, sentence comprehension, and interaction metaphor. These were addressed by McGlasha~\etal~\cite{mcglashan1996speech}. They implement an agent-based software with basic dialogue features that the user can interrogate using speech interaction.
McGlasha~\etal~\cite{mcglashan1995speech} showed that the extraction of spoken information via speech recognition effectively reduces the need for text input. The spoken language is interpreted through cascading Adaptive Speech Recognition (ASR), and the produced text is fed to the search engine. ASR is a critical component, but it can be inadequate in real-life scenarios where an extensive dictionary and the language's complexity can reduce its accuracy. Conversely, other systems have been built based on cascading ASR, such as SpeechFind~\cite{Hansen2004SPEECHFINDSD}, PodCastle~\cite{DBLP:conf/interspeech/GotoOE07}, and GAudi~\cite{4960723}.
In our research, speech recognition and language comprehension are provided by an experimenter that operates via an external software interface, to provide the user with a collection of proposed chairs (see Section~\ref{sec:sketchvoiceuserstudy}). 

\subsection{Multimodal Interaction}
\label{sec:multimodalInteraction}
Multimodal interfaces have long-standing literature. 
More than 40 years ago, the pioneering study of a voice plus gesture system was described by ``Put that there''~\cite{bolt1980put}. The user could interact with voice, gestures, and directly manipulating the graphics. The combination of speech and pointer interaction allowed the user to create a geometric figure and define its color and position. This system followed a fixed workflow where the user-created visual content and then positioned it on a 2-dimensional panel.
Since then, the combination of speech with other visual channels has been thoroughly investigated \cite{Cohen:1997:QMI:266180.266328,pentland1998smart,vo1996building,tsagarakis2006haptic,boves2002must,wahlster2006smartkom,van2009follow}, bringing attention to the advantages and disadvantages of multiple modes.
For example, Cohen~\etal~\cite{Cohen:1989:SUD:67450.67494} showed that, although gestures were suitable for model manipulation, the speech was more apt for object description, and their combination was much more effective. When designing a multimodal interface, mistakes may largely affect the final experience, as is the case of Kay's work~\etal~\cite{kay1993speech}. Here, the system uses speech and drawing interfaces, and a time-consuming vocal command control regulates the cursor's movement. 
In the last two decades, a different approach, more human-centered, has gained much attention. Human-Centered HCI (HCHCI) focuses on the person's needs as a primary aspect of completing a task. When many modalities are used, the main task here is to combine knowledge from multiple channels to achieve a coherent, continuous flow of data to be parsed. Fusing the incoming data becomes essential. Information fusion may be classified by input, design, data method, process or data source.
In multimodal HCI (MMHCI), fusion is defined as data fusion, feature fusion, or decision fusion. Data fusion occurs when data is of the same form as a multi-sensor fusion. Feature fusion, which is the most frequent approach in MMHCI, refers to tightly coupled modes such as audio or video. This type of approach involves, for example, weighted averages, Bayes estimation, Kalman filter, Hidden Markov Model~\cite{jain2005score}, neural networks. Decision fusion encompasses modalities with weak relation and is similar to cognition fusion. This fusion type can be categorized as a task-based, hierarchical, agent-based, probability-based, or component-based model. Since we use sequential audio and visual feedback (see Section~\ref{par:hybridsketchspeech}) to find our meaning, we consider our system belonging to the feature fusion category.
The combination of sketch and speech has been previously explored to accomplish a wide range of tasks. Bischel~\etal~\cite{Bischel:2009:CSS:1661445.1661670} introduced a multimodal framework to interpret the description of mechanical devices. They used two-layers neural networks to process data coming from sketch and speech inputs. 
The visual attributes extracted from the strokes define a collection of geometric features. Spatial and temporal relations are kept in consideration. Speech features are obtained with the temporal correlation between words and strokes.
In particular, sketch and speech interaction are widely explored.
Adler~\etal~\cite{Adler:2007:SSE:1384429.1384449} create an interactive whiteboard capable of processing both speech and sketch to build a shared environment. The subsequent user study allows them to use only command-based speech, annotations that replace drawings, unidirectional communication, and a limited number of visual symbols for the vocabulary.
Adler~\etal~\cite{Adler:2007:SSM:1281500.1281525} analyzed the temporal correlation between speech and sketching. They defined a set of rules for signal segmentation and alignment and used the aligned data as a basis for further studies. Our study focuses on how people combine temporal sequences of queries and how they elaborate common strategies by exploiting the advantages of sketch and voice modalities.

\section{Database and Interfaces Design}
\label{sec:databaseinterfacedesign}
We create a fine-grained dataset that would be challenging for sketch modality. Our dataset consists of 3D chairs from the ShapeNet dataset, segmented into four constituent parts, colored from a fixed selection of colors. We discuss the generation process of the dataset Variational Chairs ShapeNet (VCSNET) in Section~\ref{sec:VCSNET}. Our system combines sketch and voice modalities, leveraging the state-of-the-art 3D sketch retrieval method by Giunchi~\etal~\cite{giunchi20183d}, where a 3D sketch is depicted by the user and processed by a convolutional neural network (CNN) to create an encoding for the query. Giunchi~\etal utilize the Multi-View CNN~\cite{su15mvcnn} to create the sketch encoding, which projects the 3D sketch into multiple images around the sketch. Each view can then be encoded by the VGG-Net~\cite{simonyan2014very} and pooled to create a singular encoding for the 3D sketch. In the case of the voice input, we opt for a wizard-of-oz style experiment. We explain both of these choices in detail in section~\ref{sec:modalityInterfaceDesign} and the apparatus in section~\ref{sec:Apparatus}.

\subsection{Variational Chairs ShapeNet (VCSNET)}
\label{sec:VCSNET}
Our dataset consists of 3D chairs generated from a subset of the ShapeNet database~\cite{shapeNet}, which provides an extensive collection of 3D models with significant geometric variations. However, ShapeNet lacks two critical features: first and most importantly, it does not contain semantic tags, \textcolor{red}{which hinders text search due to the absence of attributes associated with the models. Second, it does not have variations of color and texture for fixed shapes, thus presenting low intra-model variability.}
Therefore, we generated a dataset based on ShapeNet that presents significant intra-object differences and textual meta-information, the Variational Chair ShapeNet (VCSNET) database\footnote{Variational Chair ShapeNet dataset will be made available on publication.}.
\begin{figure}
  \centering
    \includegraphics[width=\columnwidth]{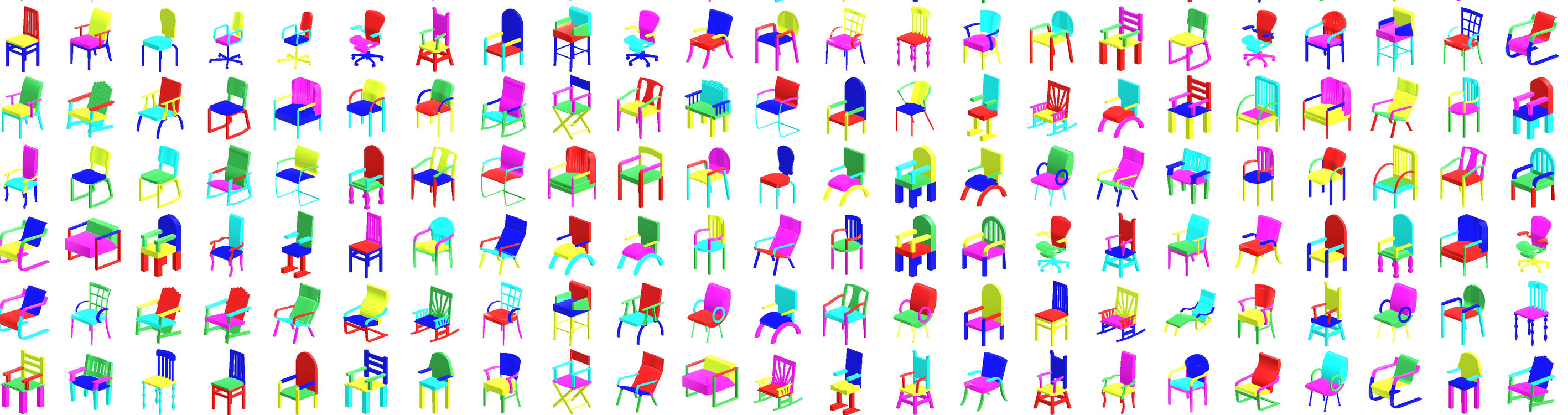}
    \caption{Our database contains $16200$ chairs generated from $45$ different shapes extracted from ShapeNet. Each shape is segmented in $4$ parts (seat, back, legs and arms). We assign colors from a fixed set of $6$ to each part following a permutation without repetitions in the same chair.}
  \label{fig:collageDBnew}
\end{figure}

To generate our database, we selected $45$ chair shapes from ShapeNet and segmented them manually with Blender into four parts (when existing): arms, back, seat, and legs. Then we generated all possible combinations of colors for each part, from a fixed set of six colors (red, green, blue, magenta, yellow, and cyan) and without having two different parts with the same color on the same chair. This permutation gives $360$ color variations and a total of $16200$ chairs in our dataset.
With visual and geometric properties, we include meta-information attached to such models to describe their semantic properties. These properties need to be descriptive and unambiguous to be effective for speech queries. This task is challenging, as the color perception by the human visual system and the ability to describe it is highly variable. Thus, we defined a fixed set of available colors, so the users do not have to select from a continuous spectrum of hue. 
Therefore, we developed a fixed dictionary of characteristics for chairs and their constituent parts. We defined a series of attributes present in all the selected chairs after analyzing their shapes. We iterated until reaching a stable number of attributes representing the elements in our dictionary (see supplemental material). 
%
\begin{figure}
  \centering
    \includegraphics[width=1.0\columnwidth]{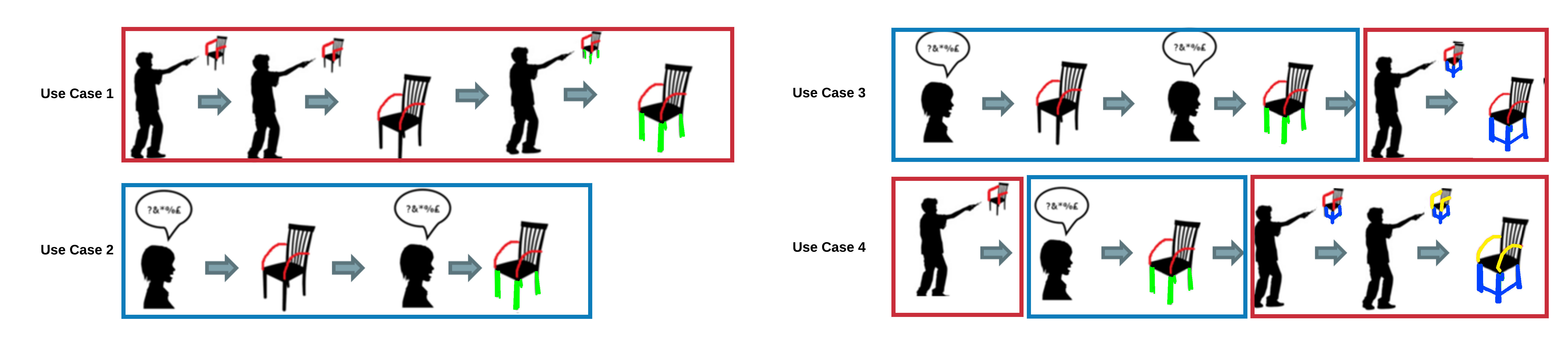}
    \caption{Plausible use cases. Red boxes contain sketch queries, blue boxes vocal queries. The first two use cases correspond to individual sketch or vocal interactions, while the last two cases combine them.}
  \label{fig:workflows}
\end{figure}

\subsection{Interface Design}
\label{sec:modalityInterfaceDesign}
\begin{figure}
  \centering
    \includegraphics[width=\columnwidth]{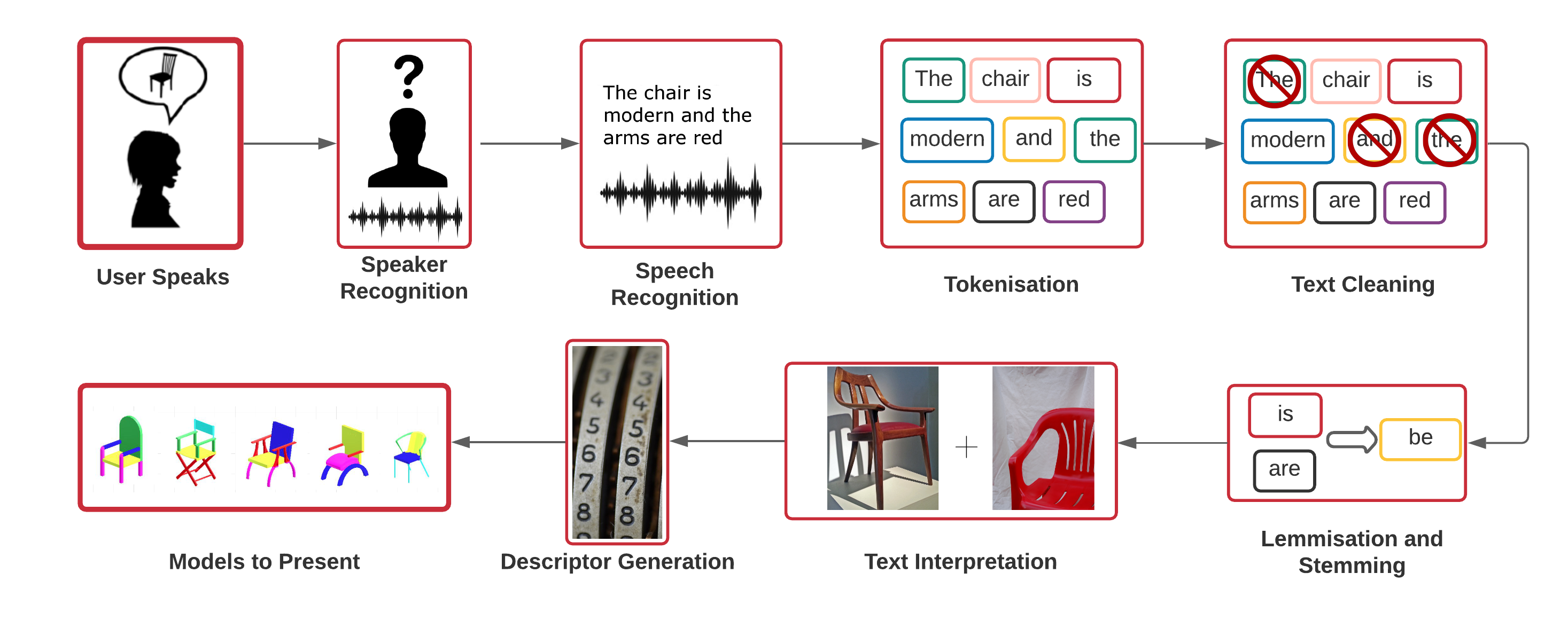}
    \caption{The speech pipeline includes: speaker identification, speech recognition, tokenization, lemmatization, stemming, text interpretation, descriptor generation, selection of the proposed results.}
  \label{fig:pipeline}
\end{figure}

\begin{figure}
  \centering
    \includegraphics[width=0.7\columnwidth]{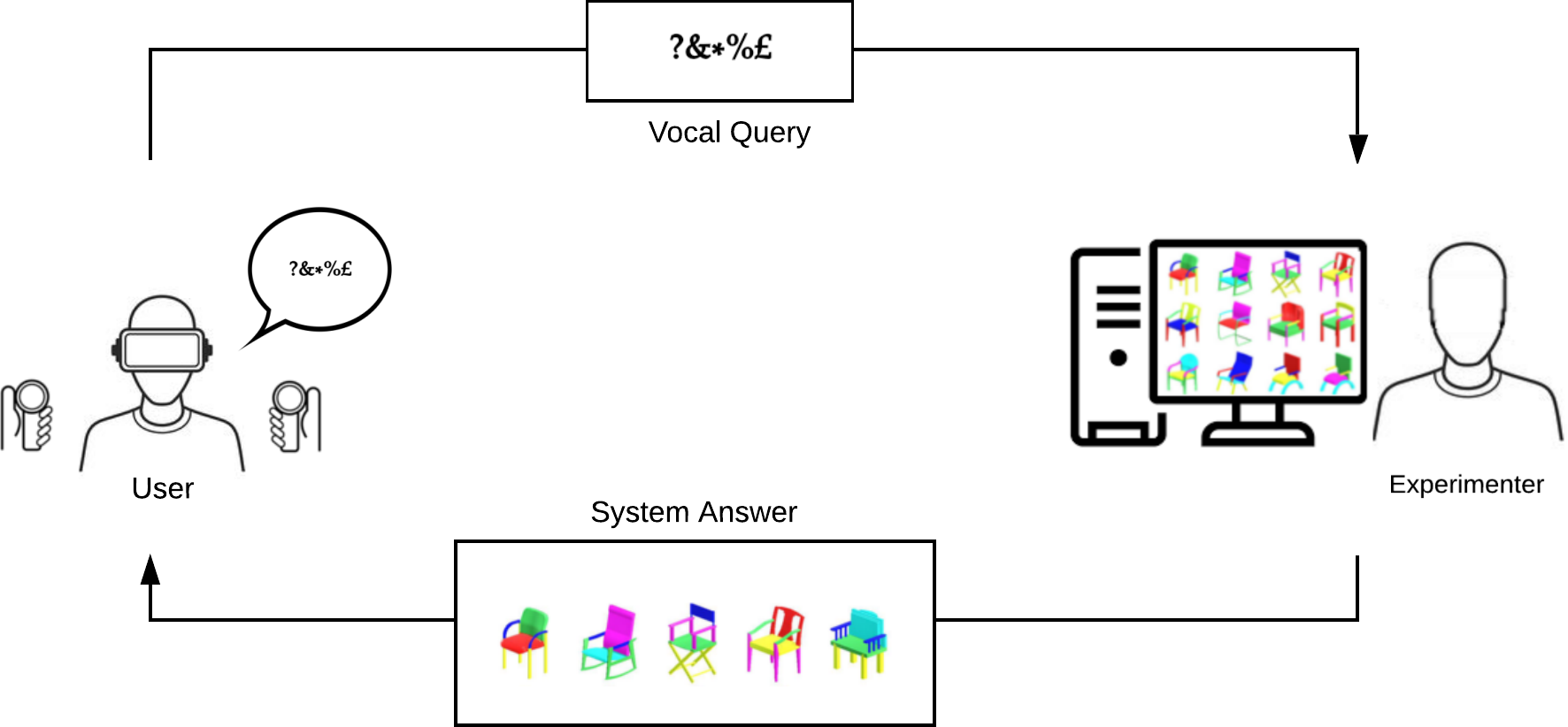}
    \caption{In the Wizard-of-Oz metaphor, the user (on the left) is immersed in a VR scene, unaware that the experimenter (on the right) handles the verbal query using the desktop interface, where he can generate the response set. The experimenter's feedback to the user is a set of five chairs displayed in a VR panel.}
  \label{fig:woz}
\end{figure}

\begin{figure}
     \centering
     \begin{subfigure}[t]{0.42\textwidth}
         \includegraphics[width=\textwidth]{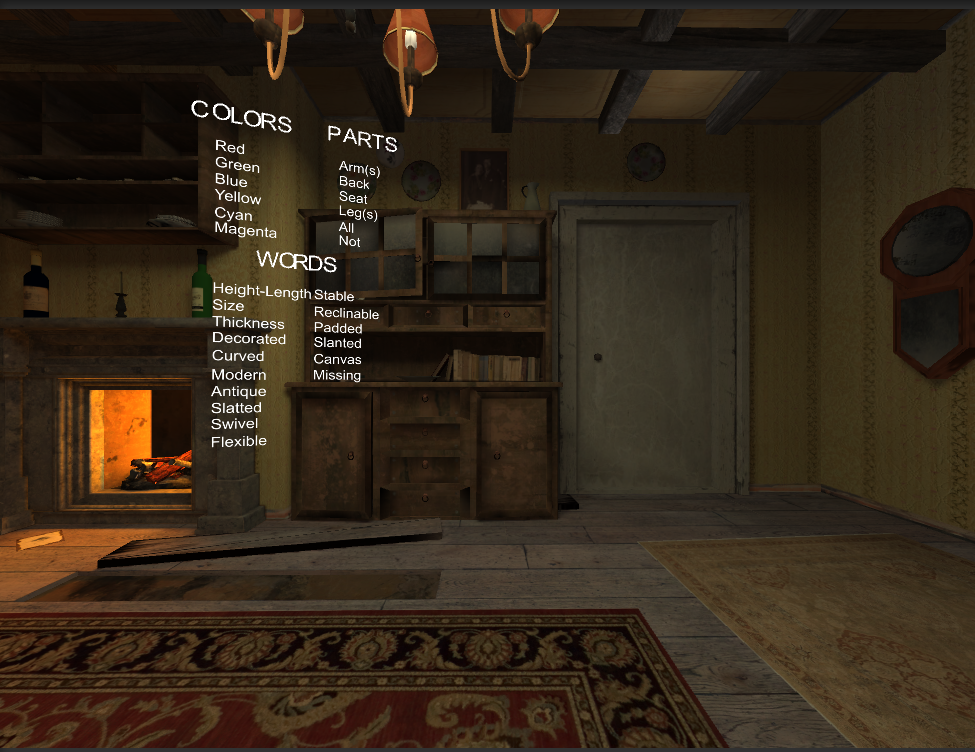}
         \caption{Each participant was immersed in this virtual environment with a panel that displays the concepts allowed for the verbal description. Synonyms and contraries were allowed.}
         \label{fig:verbalEnv}
     \end{subfigure}
     \hfill
     \begin{subfigure}[t]{0.545\textwidth}
         \includegraphics[width=\textwidth]{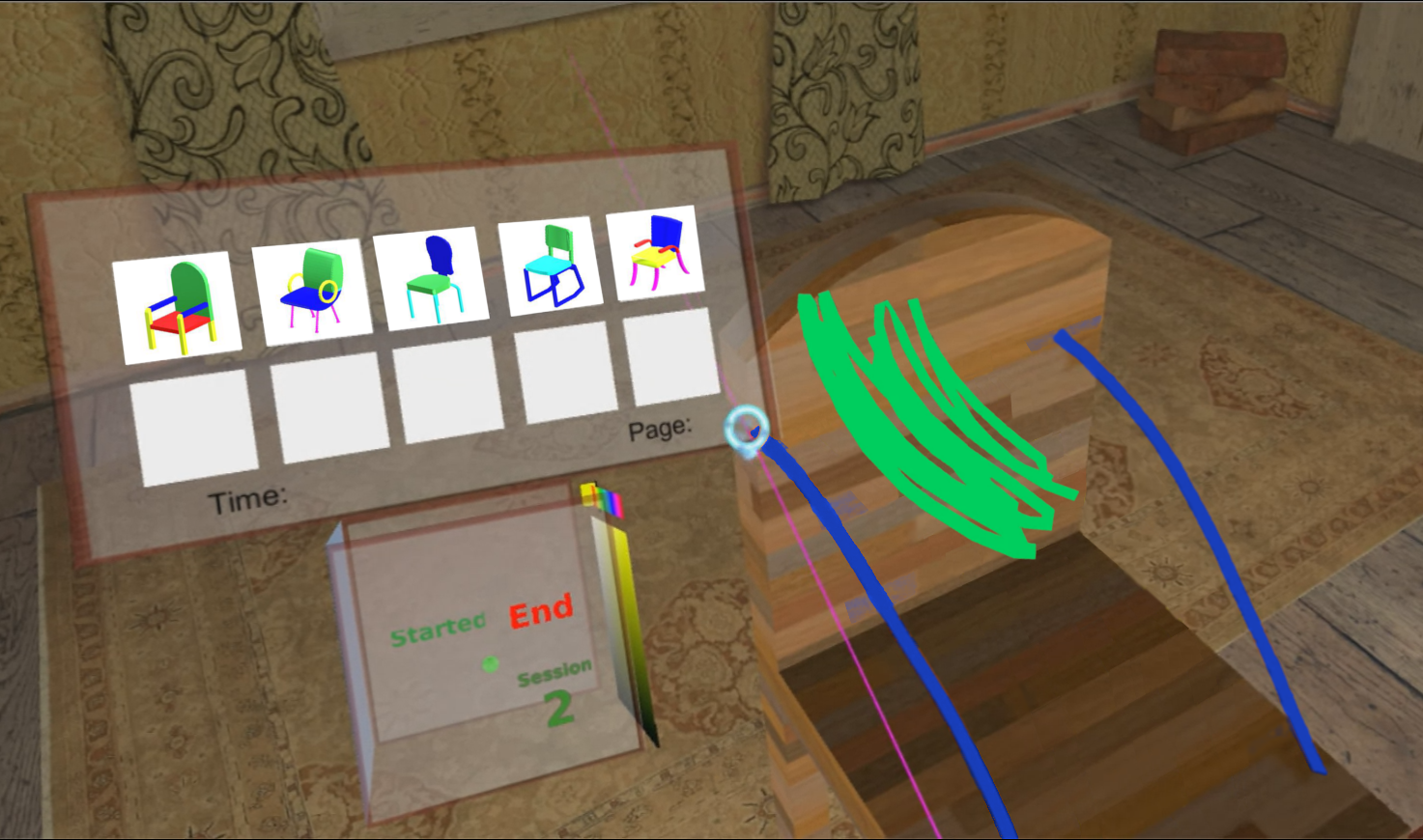}
         \caption{The sketches are done in an immersive environment, where the user can draw freely with the remote controller, and triggers the system to find the correct chair.}
         \label{fig:sketch_inter}
     \end{subfigure}
     \hfill
     \caption{VR software for speech interaction and sketch interaction.}
\end{figure}

Our system is composed of two front ends one for each modality and a hybrid search system. The user can interact with the system via voice and/or via 3D sketch while immersed in a virtual environment, in a variety of ways as illustrated in Figure~\ref{fig:workflows}. We explain the voice user interface, sketch user interface and the hybrid sketch-voice interface below:
\paragraph*{\textbf{Voice User Interface}}
A verbal description of a chair usually consists of a long sentence, ideally without hesitations or delays. The system needs to translate this description into a query that retains the information. We initially break this sentence into a set of queries, each with a subset of the total information. This technique allows the system to separate word clashes or contradictory aspects of the query. When certain problems (such as conceptual mistakes, erroneous or lossy translations, incomprehensible terms) are present in the sentence, only a partial and tiny amount of the information is dropped. To deal with common linguistic complexities (semantic ambiguities, positive or negative queries, and implicit subjects), we designed an interface that constrains the user's vocabulary.
On the other hand, an accurate interpretation of the user's speech input involves a sequence of complex steps: speaker identification, speech recognition, tokenization, lemmatization, stemming, text interpretation, descriptor generation, and selection of the proposed results, as shown in Figure~\ref{fig:pipeline}.
This pipeline can be implemented in a fully automatic or semi-automatic way. When fully-automatic, all the stages are performed by a computer while when semi-automatic, a man enters in the loop, providing the required knowledge and actions. In the second case, the experimenter operates the voice search via a \enquote{wizard of Oz} metaphor, as shown in Figure~\ref{fig:woz}. 
The experimenter considers only the relevant terms from the speech description and classifies them in the corresponding category specified in the dictionary (used as feature vector). 
\textcolor{red}{
\textit{Speech recognition} is an essential step in the pipeline that converts spoken words to plain text. During this process, relevant information must be separated from other content that can inject incorrect data into the pipeline. 
If the stage is handled automatically, dictation software can be used. 
An experimenter can convert speech to text very quickly, but its translation to written text in real-time is tough to accomplish. 
\textit{Tokenization} is a straightforward task to achieve from a corpus both if a computer or a human manages the stage. 
\textit{Text cleaning} is another phase of the pipeline that is simple to achieve. If handled by a computer, NLP libraries contain dictionaries that list all the words that do not give useful information. For a human, dropping words that do not add semantic value to the sentence is easy.
\textit{Lemmatization} and \textit{stemming} can be managed by NLP libraries and efficiently by a human with the limitations used in our experimental setting.
\textit{Text interpretation} and \textit{descriptor generation} can be managed automatically by state-of-the-art models such as the Transformer model. In this case, it is necessary to train the model, labeling all the collection with many descriptors. A possible way to achieve such a meta-information dataset is to use Amazon Mechanical Turk or hire additional participants to describe a large part of the chairs and complete the color variational dataset, replacing the description's colours. However, both options require a careful experimental design. 
As an alternative, in our experimental setup, we provide the experimenter with an interface where he can click buttons that increase the value of specific entries connected to the chair features described by the users. In this case, at the end of the query, a feature vector is automatically created and can be synchronized with the current selection in the virtual scene. 
}
Regardless of whether a human or computer software performs the voice processing, each step can introduce unexpected errors, accumulating over time. To support our choice of a Wizard-of-Oz configuration, after some tests, we noticed that speech recognition software gave us very low accuracy in some cases when managed by speech-to-text software (see supplemental material). The reasons can be found in the different components of each participant's audio profile (tempo, rhythm,  pitch, context), as well as fluency and accents, all elements that can affect the accuracy of the transcript.
Moreover, the microphone from the Oculus Rift is an additional potential source of noise in the system.
On the other side, a human can deal with speech to text conversion easily, and by limiting the domain of valid descriptive words, we achieve a reliable semi-automatic system.
It must be noted that during the entire interaction with the system, the participants are led to believe that the system is autonomous.
%
We can distinguish two interfaces that the system provides: a user interface for the participant and a user interface for the experimenter.
The user is immersed in a 3D furnished room and provided with a floating \textsc{gui} where the search results are displayed. The user verbally describes the target chair, shown on a panel positioned on the \textsc{gui}. 
The participant can use the dictionary of concepts permanently presented in front of him on a transparent layer (as shown in Figure~\ref{fig:verbalEnv}). All synonyms and antonyms are permitted, and all queries end with a terminator (we chose the word "stop").
On the other end, the experimenter, through a desktop application (as shown in Figure~\ref{fig:setupanonym}), generates a feature vector and arranges five chairs with the most similar descriptor, calculated via Euclidean distance. These chairs are sent back to the user as a system suggestion. A \textsc{gui} panel presents the chairs to the user that performs the selection, replacing the previous chair in the scene. This iterative process ends when the user is satisfied with its last selection or the elapsed time runs beyond $90$ seconds (experiment time limit).
\paragraph*{\textbf{Sketch User Interface}}
Our system reproduces the sketch interface from Giunchi~\etal~\cite{Giunchi:2018:SIM}, with minor improvements made to the snapshots generation procedure:  we switch the camera from perspective to orthogonal, and fix the camera positions and angles relative to the chair and sketches during the multi-view process.
The user can trigger the system to take snapshots of the sketch only, or the sketch plus a selected chair, chosen from the panel of $5$ query results (as shown in Figure~\ref{fig:sketch_inter}).
%

\begin{figure}
  \centering
    \includegraphics[width=0.8\columnwidth]{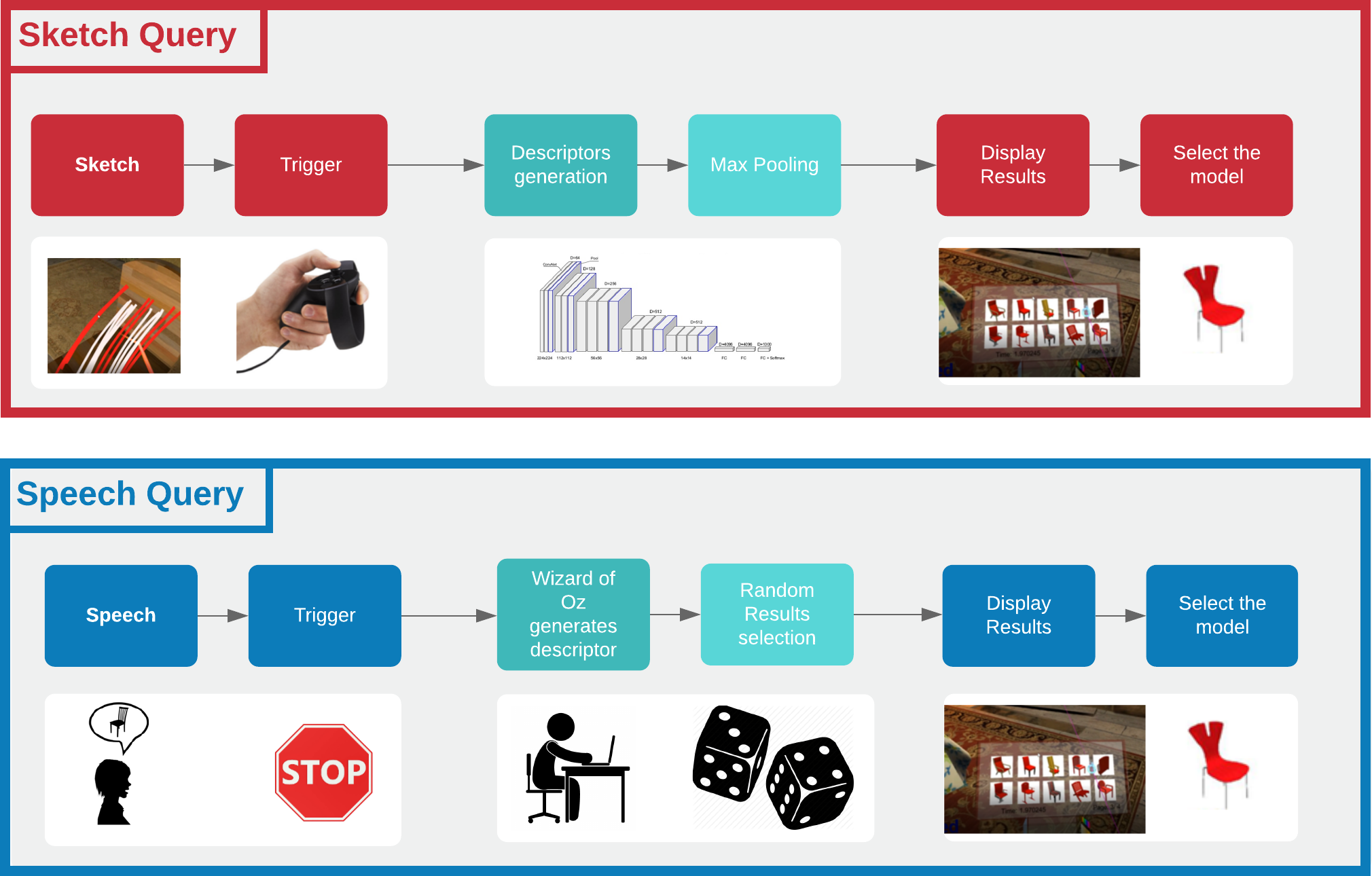}
    \caption{Sketch and speech queries contain a sequence of three phases: an interaction phase, a processing phase, and a display phase.}
  \label{fig:QueriesBlocks}
\end{figure}

\begin{figure}
    \centering
    \includegraphics[width=0.9\columnwidth]{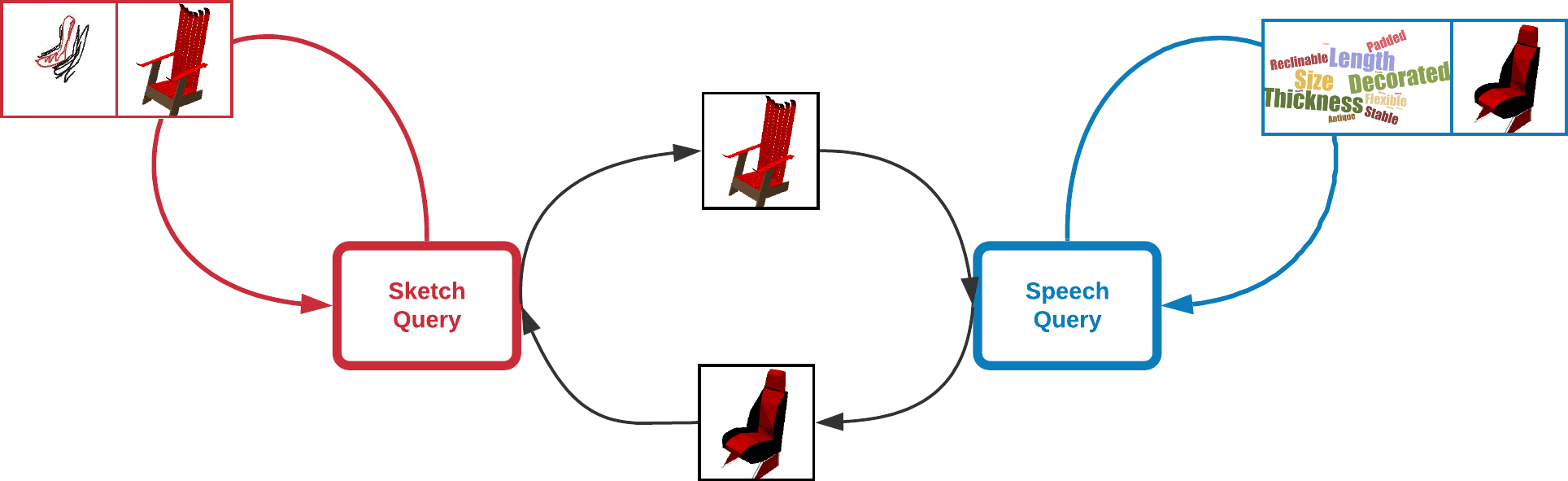}
    \caption{The sequence of queries during a search task. \emph{Red arrows}: sketch search where the user draws in 3D, and after processing the query, a final model is selected. \emph{Blue arrows}: verbal queries where the user characterizes the chair and, after a processing stage, selects a model. \emph{Center}: the connection between the queries where a selected model is the input for the next query.}
  \label{fig:Sequences}
\end{figure}

\paragraph*{\textbf{Hybrid Sketch-Voice User Interface}}
\label{par:hybridsketchspeech}
The hybrid method design, which combines speech and sketch, requires a formal definition of queries to prevent issues during the experiment. 
In our scheme, the participants are allowed to use only one method at a time, without overlapping them, but with the capacity to use any number of queries and combinations thereof. 
To avoid inconsistencies between consecutive queries of a different kind (sketch or voice), we determined that a well-formed query must complete three stages as shown in Figure~\ref{fig:QueriesBlocks}: input, processing and selection. 
In the sketch interaction (red box), the input is the set of snapshots, the CNN back-end performs the process stage, and the final stage is the selection.
For the voice interaction (blue box), input termination triggers the experimenter-in-the-loop that creates the descriptor. The selection stage is identical to the sketch query.
The query is considered well-formed if all three stages are completed, otherwise, the query is rejected. With this formalization the system can concatenate any types of query, eliminating inconsistencies and undefined chairs. 
Figure~\ref{fig:Sequences} summarizes the workflows with the combination of sketch and speech queries. We implemented the software considering these formalisms and preconditions. 

\subsection{Device}
\label{sec:Apparatus}
\begin{figure}
     \centering
     \begin{subfigure}[t]{0.53\textwidth}
         \centering
         \includegraphics[width=\textwidth]{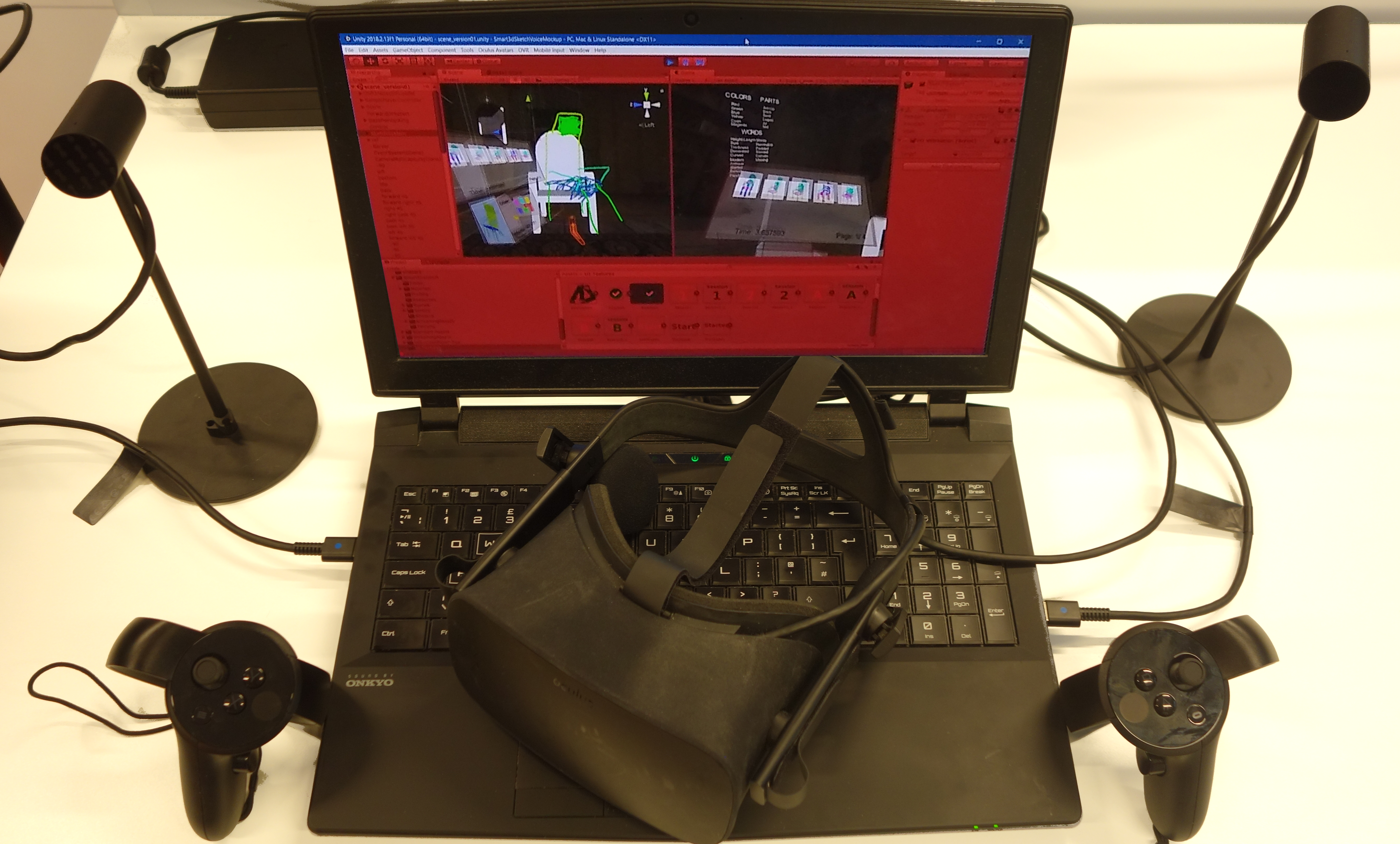}
         \caption{\label{fig:setup3}%
         Each participant used an Oculus Rift and Touch, performing all the experiments immersed in a synthetic scene. The experimenter uses the same laptop to perform the search by speech query.}
     \end{subfigure}
     \hfill
     \begin{subfigure}[t]{0.43\textwidth}
         \centering
         \includegraphics[width=\textwidth]{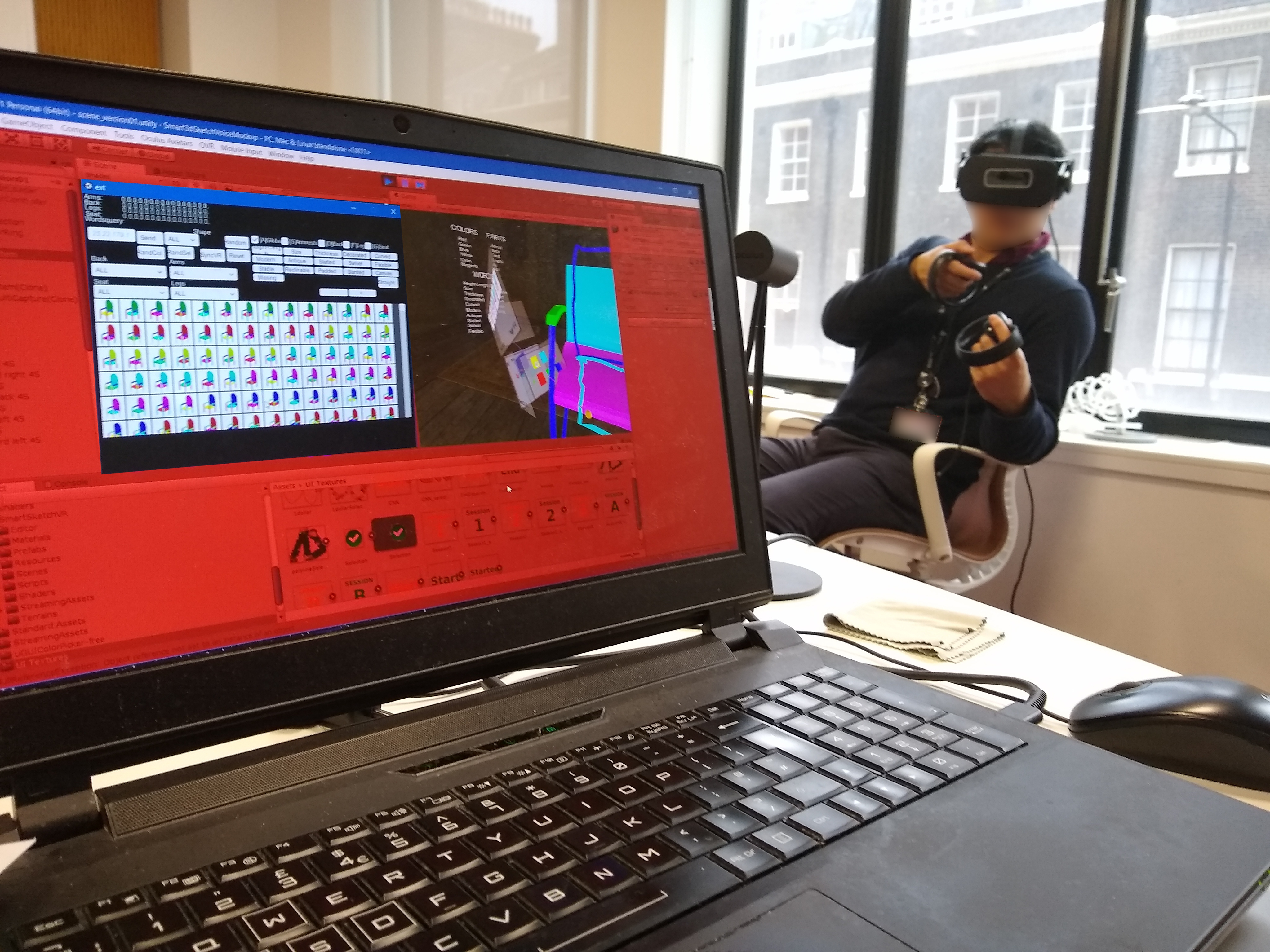}
         \caption{\label{fig:setupanonym}%
         The user interacts with our system in one of the different search modalities. The Experimenter-on-the-loop uses an optimized GUI to quickly display the chairs resulting from the verbal queries.}
     \end{subfigure}
     \caption{Setup used for the experiments.}
     \hfill
\end{figure}

Our experimental setup consists of an Oculus Consumer Version 1 (CV1) head-mounted display (HMD) paired with Oculus Touch controllers. The experiments were performed using a PC laptop with an Intel i7-6700 CPU Processor, mounting an NVIDIA GeForce GTX 980M graphics card and 64 GB of RAM (see Figure~\ref{fig:setup3}). The experiment-in-the-loop interface runs in the same laptop and is connected to VR application with a socket. The VR application and the 2D desktop software for the experimenter-in-the-loop were developed with Unity3D (v. 2018.2.13f1), as shown in Figure~\ref{fig:setupanonym}.


\section{First User Study: Speech interaction query optimization}
\label{sec:study1}

We designed two user studies to evaluate our multimodal solution, approved by the \textbf{ANON} ethics board. The purpose of the first study was to determine the optimal number of words-per-query in our experiment settings, given that the speech input needs to be interpreted by a human experimenter. \textcolor{red}{We balance speech query length between the expressiveness of the user and the recall ability of the experimenter, which was the same for both studies.} We measured the success rate of 3D model retrieval within our database. We computed the optimal number of words-per-query in our first study and used it as a fixed number in the second experiment, detailed in Section~\ref{sec:study2}.
\subsection{Voice Interaction for 3D Object Retrieval}
\label{sec:voiceuserstudy}
In the first user study, the participant was seated and was asked to find a chair from a digital collection by verbally describing the proposed target chair in the English language, while immersed in VR. The user was placed in a virtual room with a chair that acted as a placeholder, and a panel with the list of chair features allowed for speech input (Figure~\ref{fig:verbalEnv}). 
This initial chair was replaced when a user selection was made. The target chair was shown in an image positioned in a \textsc{gui} attached to the user's left wrist.
We asked the participants to use a sentence that described one or more features of the chair. Given the nature of the vocal query, we explained that the system could not manage very long sentences, and thus each spoken query should not be longer than $10$ seconds. The user had $90$ seconds to complete the search. We reduced the complexity using a dictionary with $30$ chair attributes, allowing synonyms and antonyms. 
We hired ten university students ($7$ males, $3$ females, between $18$ and $43$ years old), to participate in our study. Each participant performed $27$ search sessions. We divided these sessions into three groups of nine randomized models, where the same n-gram (number of words per query) was used. We also randomized the order in which these n-gram modalities appeared to the participants. The participants were unaware of the experiment's goal, which was to evaluate the proficiency of three different n-gram modes: bi-grams, 4-grams and 6-grams. Beyond the accuracy and the time to complete the search, we tracked the job attempts. During the entire process, the users were made to believe that the system was fully autonomous.
Before starting the test, we asked the users to fill a form to record previous experiences with VEs, 3D software, or games. \textcolor{red}{Via a 5-points Likert scale, we asked them to self-rate their experience with VR (mean $2.3$) and self-evaluate orientation (mean $2.7$), perspective-taking (mean $2.8$), way-finding (mean $2.8$), and visualization of 3D objects (mean $2.9$).} At the end of the experiment, we recorded the number of successes and the time needed to complete the task. As the last step, the participants were asked to evaluate the system's proficiency and propose possible improvements to the application, such as better response time, different words to describe the chairs or a more intuitive interface. They were additionally asked to score from one to ten, whether they perceived the system as robotic or human.
Once the first set of experiments was concluded, an optimal number of n-grams was determined, which was later used for the second set of user experiments (Section~\ref{sec:study2}).
%

\subsection{First User Study: Results}
\label{sec:results1}

We analyzed the differences and outcomes of the n-gram sessions described in Section~\ref{sec:voiceuserstudy}. The precision was measured by the number of times the user found the target chair. \textcolor{red}{Across different sessions, only the number of n-grams and the target models changed.}
We summarize the results from this experiment in Table~\ref{table:experiment1}. We show that the 6-gram mode achieved the best results. After completing the experiment, we asked the users to evaluate the system's performance, rating it on a Likert scale from $1$ (very bad) to $5$ (very good), resulting in an average score of $2.6$.
\begin{table}[ht!]
\centering
\begin{tabular}{@{}ccccl@{}}
\toprule
        & Precision           & Average time (s) & Average \#queries \\ \midrule
2-grams & $33.3\%$ & $81.3$           & $6.28$ \\
4-grams & $34.4\%$ & $78.2$           & $5.21$ \\
6-grams & $\mathbf{43.0\%}$  & $\mathbf{75.5}$  & $\mathbf{5.03}$ \\ \bottomrule
\end{tabular}
\vspace{0.5em}
\caption{Results from the first study showing an optimal number of 6 words per query.}
\label{table:experiment1}
\vspace{-4mm}
\end{table}

Before running the study, we ran a pilot study to evaluate queries of more than six components; however, this quickly became very difficult to manage for the experimenter. We avoided testing single-word queries because we observed that a noun is always paired with a variable number of adjectives. 
The results showed that six-gram was the right choice between the tested modes, which simultaneously was the limit we took into account during the experiment. 
\textcolor{red}{Moreover, unlike in the case of 2- and 4-grams, we did not register repetitions of attributes in consecutive 6-grams queries.}
\textcolor{red}{
Additionally, we noticed that a limited palette helped the users to make quick and accurate descriptions of the colors. For this purpose, a simple 2-gram is sufficient, and users normally concatenate different parts and related colors. On the other hand, the description of the shape is more challenging, and users tended to use more n-grams to characterize the global style of the chair or its component.
}
The first study results and the gained experience allowed us to implement multiple improvements for the second study: we modified the interface used by the experimenter for feature-vector generation, adding shortcuts such as a reset button, and buttons for increasing and decreasing an attribute's value. We added a user preparation phase before the experiment, where we clarified to the participant the meaning of all chair attributes, showing graphical examples of real chairs. This preparation was required to level the field between native and non-native English speakers hired for the experiment. We included a self-evaluation of English level in the questionnaire after the study, on a Likert scale from $1$ (very bad) to $5$ (very good). As some users noticed that the system was not fully automatic because of the mouse buttons' sounds, we replaced it with a noiseless trackpad.
\subsection{Experimental adjustments from user feedback}
The users involved in the first study provided useful suggestions on how to improve the quality and the proficiency of the voice interface. Following their suggestions, we cut down the dictionary from $30$ to $20$ concepts. We additionally removed words not used during the study or that led to ambiguities. As an example, we combined attributes that represented similar concepts such as ``curvy'' and ``wavy''. Furthermore, we introduced additional terms indicated by the users that were not present in the first version of the dictionary, and we set up discrete steps for the change of feature values. In this way, we avoided in-between values, overflows, or under-flows associated with each feature in the descriptor.
We framed characteristics related to the size, length and thickness to prevent mistakes related to spatial dimensions. 
To reduce the latency in feedback introduced by the experimenter operating the Wizard-of-Oz interface,
we added shortcuts to the 2D desktop interface. We implemented a way to synchronize the current chair located in the room with the existing descriptor. This prevented the issue of losing features between two consecutive queries. Moreover, we improved the dictionary layer's visualization, where the concepts were displayed, rearranging the list, and positioning it closer to the center of the frustum.
\textcolor{red}{
The outcomes of the first experiment did not lead to modifications of the sketch interface. We designed the queries and their generation as separate interactions that work on different solution spaces, but simultaneously compatible in our search pipeline.}

\section{Second User Study: Sketch and Speech for model retrieval}
\label{sec:study2}
In this section, we describe and discuss the second experiment, where we evaluate three different modes of interaction: speech interaction, sketch interaction, and both interactions combined. For the speech interaction, we use a fixed number of words-per-query obtained from the first user study.
\subsection{Second User Study: Sketch and Voice Interaction for 3D Object Retrieval}
\label{sec:sketchvoiceuserstudy}

In the second study, the user was immersed in the same environment of the first experiment. This study compared the different techniques for 3D object retrieval in a large collection of detailed models. Each user performed three sessions: sketch interaction, speech interaction and the combination of sketch and speech.
Similar to our previous study, here, the three interaction modes appeared in random order. We used the same furnished room in VR with a default chair as the starting point, and we enabled or disabled the interactions according to the session type. The speech interaction method was the same as the first study, with the only exception that in the second study, we utilized the fixed number of n-grams that was determined to be optimal in the first study.
The sketch interaction, based on the work by Giunchi~\etal~\cite{Giunchi:2018:SIM}, provides the user with a visual metaphor to search the item. The sketches consist of colored stripes and the color is selected using a simplified palette. The user draws the 3D sketch in the scene and then triggers the system to retrieve the best chair, as shown in Figure~\ref{fig:sketch_inter}.
The system proposes a set of chairs that the user can navigate and select one. Every time a visual query is requested, $12$ snapshots of the sketch are taken from different fixed angles and fed to the neural network for evaluation. While Giunchi~\etal implemented a palette that allows a continuous selection of hue and saturation, we limited the palette to six colors present in the database. The user can sketch and trigger the system with a pure sketch input or draw on top of the selected model and generate a descriptor that is a combination of the sketch and the selected chair model.
Ten new participants were hired to participate in this second experiment. As in the first experiment, they were undergraduate and postgraduate students from different university departments. However, there was no overlapping with the participants from the first study. Five of the participants were male ($5$ female), and the participants' ages varied from $18$ to $43$ years old. Five of the participants either had no previous experience with VR or had tested it once in their lives. Each user carried out the retrieval of $27$ models, divided into groups of $9$ models for each type of session (sketch, voice, and sketch + voice) in random order. Each retrieval query lasted for a maximum of $90$ seconds. We recorded the success rate, the time needed to complete the task, and through a final form, we asked the users to self-evaluate their English level and score the user experience.
\subsection{Second User Study: Discussion}
\label{sec:results2}
\begin{figure}
     \centering
     \begin{subfigure}[t]{0.48\textwidth}
         \includegraphics[width=\textwidth]{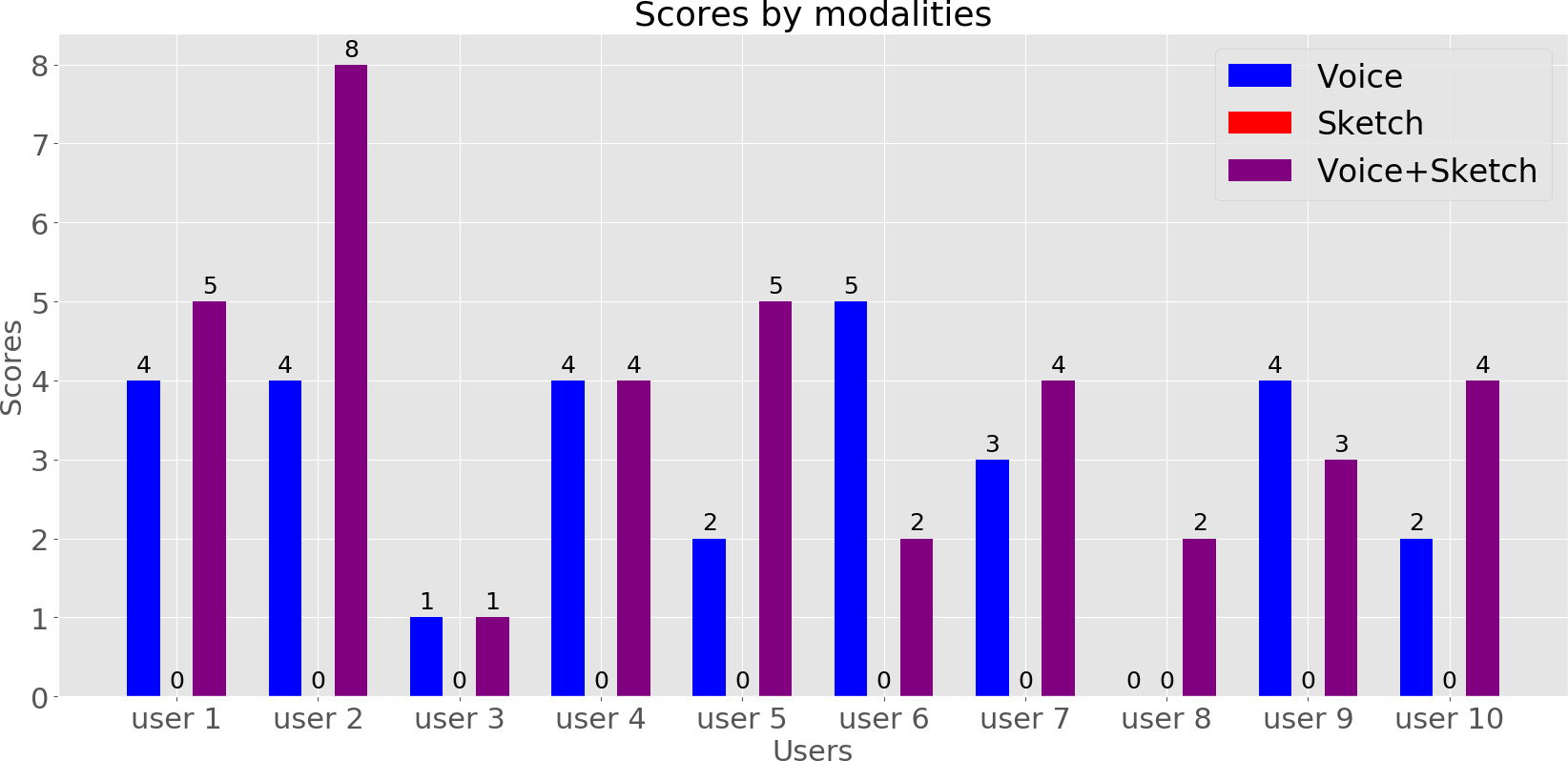}
         \caption{%
         }
         \label{fig:targetSuccesses}
     \end{subfigure}
     \hfill
     \begin{subfigure}[t]{0.48\textwidth}
         \includegraphics[width=\textwidth]{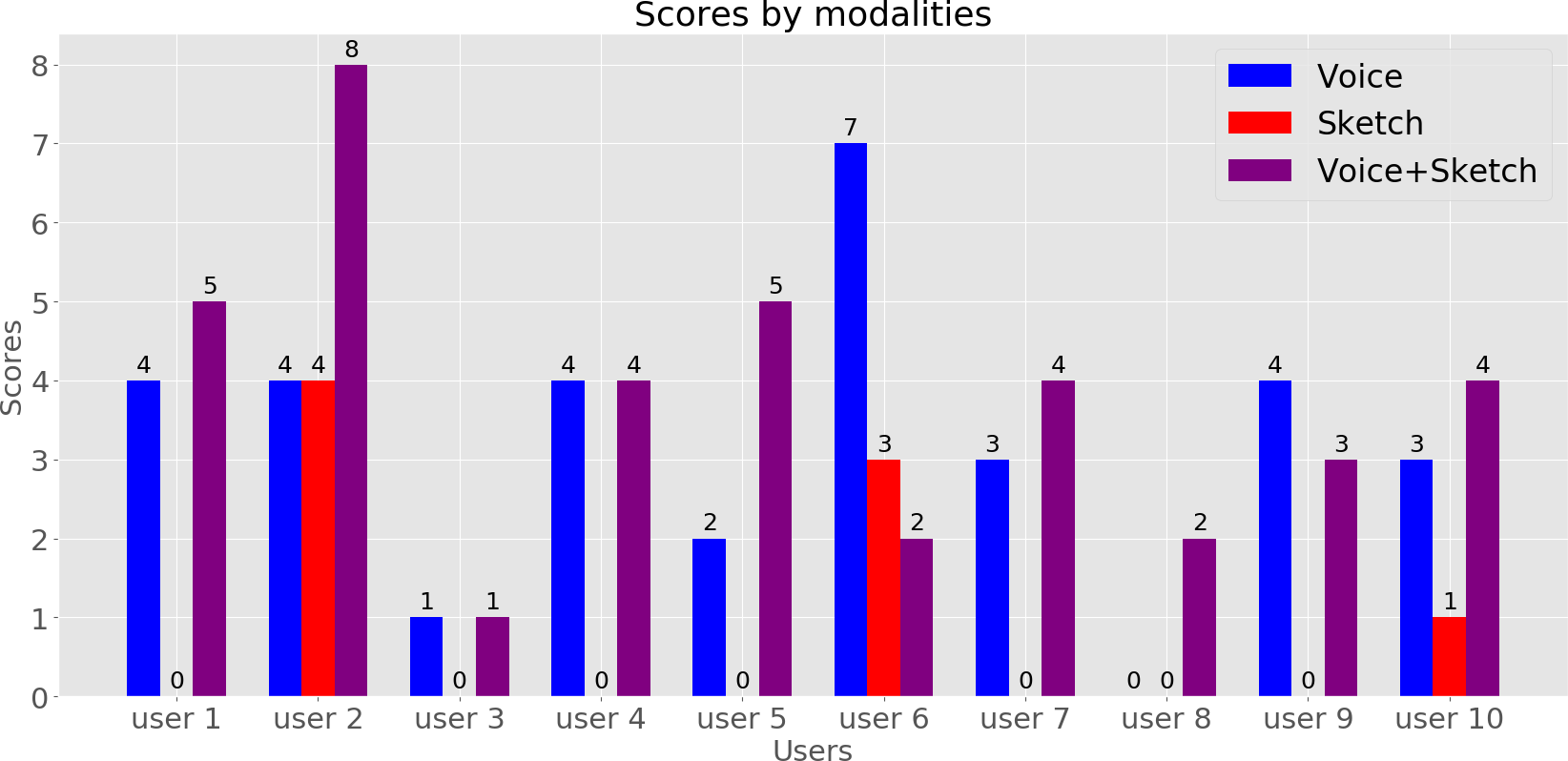}
         \caption{%
         }
         \label{fig:shapeSuccess}
     \end{subfigure}
     \caption{(a) The number of successfully found chairs for each user, for the three modes of interaction. As reflected by the results, our database is specifically designed to be challenging to navigate by sketch retrieval. This is caused mainly by the system's difficulty to pair the correct color to the exact part of the chair. (b) The number of successfully found shapes (Y-axis) vs User ID, for the three modes of interaction. For some users, this shows an increase in precision for the voice only and sketch only interactions.}
     \hfill
     \label{fig:targetAndShapeSuccess}
\end{figure}

We show in Figure~\ref{fig:targetAndShapeSuccess} the number of successes for each user, and Table~\ref{table:experiment2} the comparative modality results of the second experiment.
While it can be seen that sketch achieved a very low precision, it was expected by the design of the database, which for sketch, creates a high degree of ambiguity in the correspondence between colors and chair parts (see sec.\ref{sec:databaseinterfacedesign}). Thus, we further studied the retrieval statistics but considering only the chair's shape as the target independently of color. 
In this case, we found that voice interaction and sketch alone improved their precision, while the hybrid interaction's precision remained unchanged (see Figure~\ref{fig:shapeSuccess}).
We note that the voice query is very effective with colors and a valid method to describe the chair's shape. However, not all the shapes are easy to describe or disambiguate from other shapes with similar features. 
\begin{table}[ht!]
\centering
\begin{tabular}{@{}lrccccl@{}}
\toprule
             & Precision         & Precision (shape)    & Avg. time (s)  & Avg. \#queries & User score \\
\midrule
Voice        & $32.2\%$          & $35.5\%$                  & $\mathbf{80.5}$   & $\mathbf{3.5}$    & $\mathbf{3.7}$ \\
Sketch       & $0.0\%$           & $8.88\%$                  & $90.0$            & $2.7$ & $2.4$ \\
Sketch+Voice & $\mathbf{42.2\%}$ & $\mathbf{42.4\%}$         & $81.4$            & $3.4$             & $3.6$ \\
\bottomrule
\end{tabular}
\vspace{0.5em}
\caption{Results from the second experiment, comparing the three modes of interaction for 3D object retrieval.}
\label{table:experiment2}
\vspace{-4mm}
\end{table}
Two circumstances were identified where the user could not progress in the search. The first occurred when the results shown to the user did not change with respect to the last query. The second case happened when the system constantly displayed chairs that shared similar attributes with the target without showing the correct one. 
In most situations, if the user managed to isolate the correct shape using voice queries, the right combination of shape and color was also found. Indeed, the majority of remaining difference between the target and the current chair was the color, that could be addressed with one or at most two queries. In particular, the results from the sketch did not have colors coupled with the right component, even when the shape was found. 

\begin{figure}
    \centering
    \includegraphics[width=0.70\columnwidth]{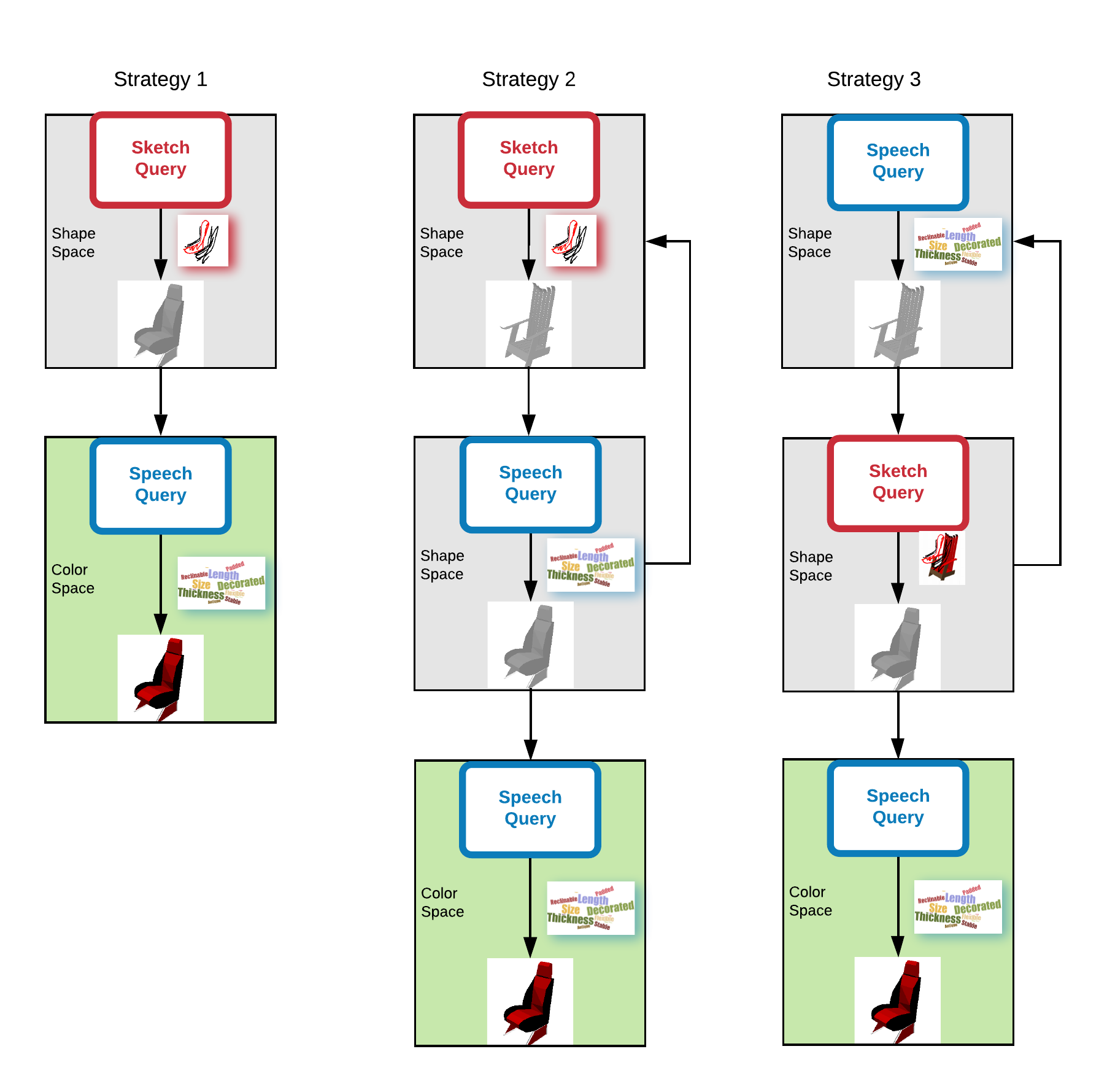}
    \caption{Flow-charts of common strategies in hybrid interaction. \emph{Left}: sketch query for the target shape, followed by a vocal query for the colors. \emph{Middle}: iteration between sketch and speech queries to search for the target shape, ending with the vocal to set the colors. \emph{Right}: start with a vocal query to find the shape and iterate with the sketch, ending with speech for colors.
    }
  \label{fig:strategiesSketchVoice}
\end{figure}

Several strategies emerged during the hybrid interaction sessions (Figure~\ref{fig:strategiesSketchVoice}). Some users began looking for the exact shape using voice alone interaction, without considering the color. When the query was terminated with a selected chair, the descriptor was definite and paired with that last selection. Thus, there was a common feature vector on the next five chairs. With the next query, an incremental description was added, which was focused on the shape or color, refining the search.
When the shape was found, a maximum of two queries described all the colors present in the chair.
During the sketch interaction, the user selected the correct colors and drew a complete sketch or continuously tried to trigger the search with the chosen model and additional sketch. In the second case, the user tried to complete the missing parts of the chair or replaced the part if the color did not match. We observed that even when the shape was successfully retrieved, the colors were usually wrong. The sketch search used the same colors depicted by the user, but often does not associate it with the right part of the chair.
The most successful outcomes were obtained through a sequence where both voice and sketch queries were utilized. From the strategies we identified three primary query strategies that provided good results, with small differences or additional loops. The first occurs when the user sketches to find the shape and speaks to find the color. The second is sketching to find the shape and select a non-target chair. Then, the user begins a new search with vocal interaction, refining the search. 
When the shape is correct, the user selects colors with voice. The third strategy starts with speech to find the right shape and then sketch to draw details on the chair selected to replace the current one in the scene. Finally, the user uses the voice to determine colors.
The voice's proficiency over the sketch occurs because we can verbally describe some features properly without interpretation error. \textcolor{red}{While the dataset was selected for high shape variability, increasing its complexity (e.g. adding texture to shapes) could further limit the voice modality's effectiveness, and make it more difficult for the experimenter to interpret.}
On the other hand, combining sketch and voice offers multiple advantages. 
Firstly, it does not change focus, and the user can operate independently of his hands' location. Second, the context switching for the user is more immediate, compared with other hands or interface interactions; and thirdly, the user can explore a completely different set of features created by the input modality.
\textcolor{red}{
3D models in the dataset store information in two different ways: text and geometry with colors.
Verbal descriptions present some advantages for the user over sketching, especially if utilizing a limited dictionary of characteristics. Query composition is faster than sketching in 3D, and it can be quite accurate if the right words are selected, which was often exploited by users during our studies.
In addition, we noticed that visual appearance is important for both modalities, while text information is only relevant for speech queries. 
In order to formulate a query, users were able to rely on stored knowledge from past queries, by simply observing the model's visual appearance. On the other hand, implicit text descriptors are not accessible or known by the user, and are only used to compute comparisons of vectors generated by speech interaction.
}
Finding the shape proved to be the most difficult task as the description is subjective and ambiguous for some dataset items. Combining two modalities tends to improve the results with respect to using a single one.
Voice and sketch present non-overlapping limitations, hence with their combination, the user can overcome pitfalls by selecting the best query.
We conducted a Friedman test over our results considering the methods as independent variables and rate of success as a dependent variable, using pairwise comparisons between methods (SPSS Statistics, 2019). The pairwise comparisons (p-value 0.05) show that 3D sketch is the worst method, while the other two methods can not be considered different. While this analysis agrees with the comparison of modalities in Figure~\ref{fig:targetAndShapeSuccess}, we would argue that it does not consider the iterative search and the complementary nature of the sketch modality within a sequence of searches. The benefit can be seen by the different search strategies that emerged, and that sketch was not disregarded as ineffective by the participants.
As in the previous experiment, we asked the participants to evaluate through a questionnaire their user experience for the three modalities using a Likert scale with values between $1$ (very bad) and $5$ (very good). Vocal interaction was rated with an average of $3.7$ out of $5$, sketch scored $2.4$, and sketch with voice $3.6$.
Also, we asked the participants to evaluate whether the retrieval system was perceived as fully automatic or not. The range of values varied from $1$ (fully managed by a human) to $5$ (fully automatic), and we obtained an average of $2.9$.

\section{Limitations and Future Work}
We explored a novel interface that combines sketch and speech to describe and retrieve 3D objects from a variational color database.
\textcolor{red}{Such database was achieved by segmenting a subset of ShapeNet chairs, but other collections may have their own variation that makes them suitable (such as PartNet~\cite{DBLP:conf/cvpr/YuLZZ019}).}
We showed that sketch and voice interaction can be used together efficiently, however, they also fail independently in different scenarios.
As this is an exploratory study and our interface does not contain complex functionalities (e.g. gestures), we plan to extend both the number of participants and to improve the VR software with additional components.
\textcolor{red}{
By analyzing the emerging strategies in our user studies, we noticed that sketch interface could be improved by separating the shape component from color. This could be done by implementing a new interface similar to ShadowDraw~\cite{DBLP:journals/tog/LeeZC11}, where only the shape is considered during the search.
}
In the future, we aim to replace parts of the experimenter-in-the-loop tasks with automatic processes controlling possible error injection and evaluating their impact on the speech interaction pipeline.
In particular, speech recognition can be replaced by software possibly after a quick training. We can then apply classic Natural Language Processing (NLP) techniques to identify meaningful words from the speech input, and replace human text interpretation with the state-of-the-art neural network for NLP~\cite{2020arXiv201010504Z}. In this case, we also need to associate additional information to the models. The resulting dataset should contain model descriptions of shape and color, improving on our naive solution that relies on a dictionary.

\section{Conclusion}
We have studied the effect of combining sketch and speech within a virtual environment for 3D model retrieval. To achieve this, we designed a new dataset that is challenging for each modality independently, by introducing color variations on each of the parts of the original chairs from ShapeNet. In this dataset, we showed that the sketch interaction has difficulties recognizing the correct pairing between colors and specific components of the model, as semantic information makes sketch search ineffective on complex objects. At the same time, by using vocal descriptions, it is challenging to express the overall appearance of the object, demonstrated by the users combining using sketch queries to convey this information. Even if the sketch modality alone achieved poor performance, this combined process marginally reduced the time to find the target.  
The implementing multimodal system users were able to interact through both sketch and voice, therefore providing two independent techniques to explore feature spaces of shape and color. Our design allows the consistent interchange of feature descriptors from both interactions, enabling the integration of speech and sketch queries in a single search session.
We conducted two user studies. The first was a preliminary study to determine optimal parameters for the speech interaction. The second experiment compared sketch interaction, voice interaction, and a combination of both in the context of 3D model retrieval in an immersive environment.
During our studies, we observed the emergence of common new search strategies, enabled by the seamless integration between interactions in our system. We analyzed these strategies and discussed possible improvements to the interface. Finally, we analyzed the different modalities' performance in terms of accuracy, speed and user experience rating. Our results show the poor performance of the sketch-only interaction in the VSCNET database. Results suggest an improvement of the multimodal interface with respect to both separate modalities, although with insufficient statistic to definitively establish the lead with respect to voice interaction. This could be explained by the difficulty in disentangling the benefits of sketch in the hybrid method with the proposed system's iterative search.


\bibliographystyle{ACM-Reference-Format}
\bibliography{main}

\end{document}